\RequirePackage{fix-cm}
\documentclass[smallextended]{svjour3}       
\smartqed  
\usepackage{graphicx}

\usepackage[sort&compress]{natbib}
\bibpunct{(}{)}{;}{a}{}{,}
\usepackage{hyperref}
\usepackage{amsfonts}
\usepackage{amsmath}
\usepackage{setspace}
\usepackage{longtable}
\usepackage{graphicx,verbatim,bm,color}
\hypersetup{colorlinks=true, citecolor=blue, linkcolor=blue, urlcolor=blue, filecolor=black}

 \journalname{TEST}
\begin{document}

\title{Discussion of {\it Single and Two-Stage Cross-Sectional and Time Series Benchmarking Procedures for SAE}
}

\titlerunning{Discussion of Pfeffermann et al. (2014)}        

\author{
        Rebecca C. Steorts  \and
        M. Dolores Ugarte 
}

\authorrunning{Discussion of  \cite{pfeffermann_2014}  } 

\institute{
 Rebecca C. Steorts \at
             Department of Statistics, Carnegie Mellon University\\ Baker Hall 132, Pittsburgh, PA 15213\\
                \email{beka@cmu.edu}
         \and
           M. Dolores Ugarte  \at
             Department of Statistics and O.R., Public University of Navarre\\ Campus de Arrosadia, 31006 Pamplona\\
                \email{lola@unavarra.es}
}

\date{Received: date / Accepted: date}
\maketitle

We congratulate the authors for a stimulating and valuable manuscript,
providing a careful review of the state-of the-art in cross-sectional and
time-series benchmarking procedures for small area estimation. They develop a
novel two-stage benchmarking method for hierarchical time series models, where
they evaluate their procedure by estimating monthly total unemployment using
data from the U.S. Census Bureau. We discuss three
  topics: linearity and model misspecification, computational complexity and
  model comparisons, and, some aspects on small area estimation in
  practice. More specifically, we pose the following questions to the authors, that they may
  wish to answer:
   How robust is their model to misspecification? Is it time to perhaps move away
  from linear models of the type considered by \citep{fay_1979,battese_1988}?
What is the asymptotic computational complexity and what comparisons can be made to other models?
Should the benchmarking constraints be inherently fixed or should they be random?

\section{Linearity and Model Misspecification}
\label{sec:linear}

The authors review previous work in cross-sectional and time-series
benchmarking thoroughly, and propose a new cross-sectional
two-stage time series benchmarking method.  This method is mathematically
well-posed and carefully thought-through. It is applied to the estimation of unemployment
totals at the state-level and census-divisions (aggregate state-level
estimates).  However, with respect to model validity, the analysis does not go
beyond evaluating coefficients of variation with respect to the direct
estimates. Furthermore, although the application here \emph{may}
satisfy the modeling assumptions, it is important that
methods are built to be robust to general model misspecification.
As one of the authors recently writes, checking model validity is very important in small area estimation \citep{pfeffermann_2013}.\\

\begin{figure}[htbp]
\begin{center}
\includegraphics[width=0.8\textwidth]{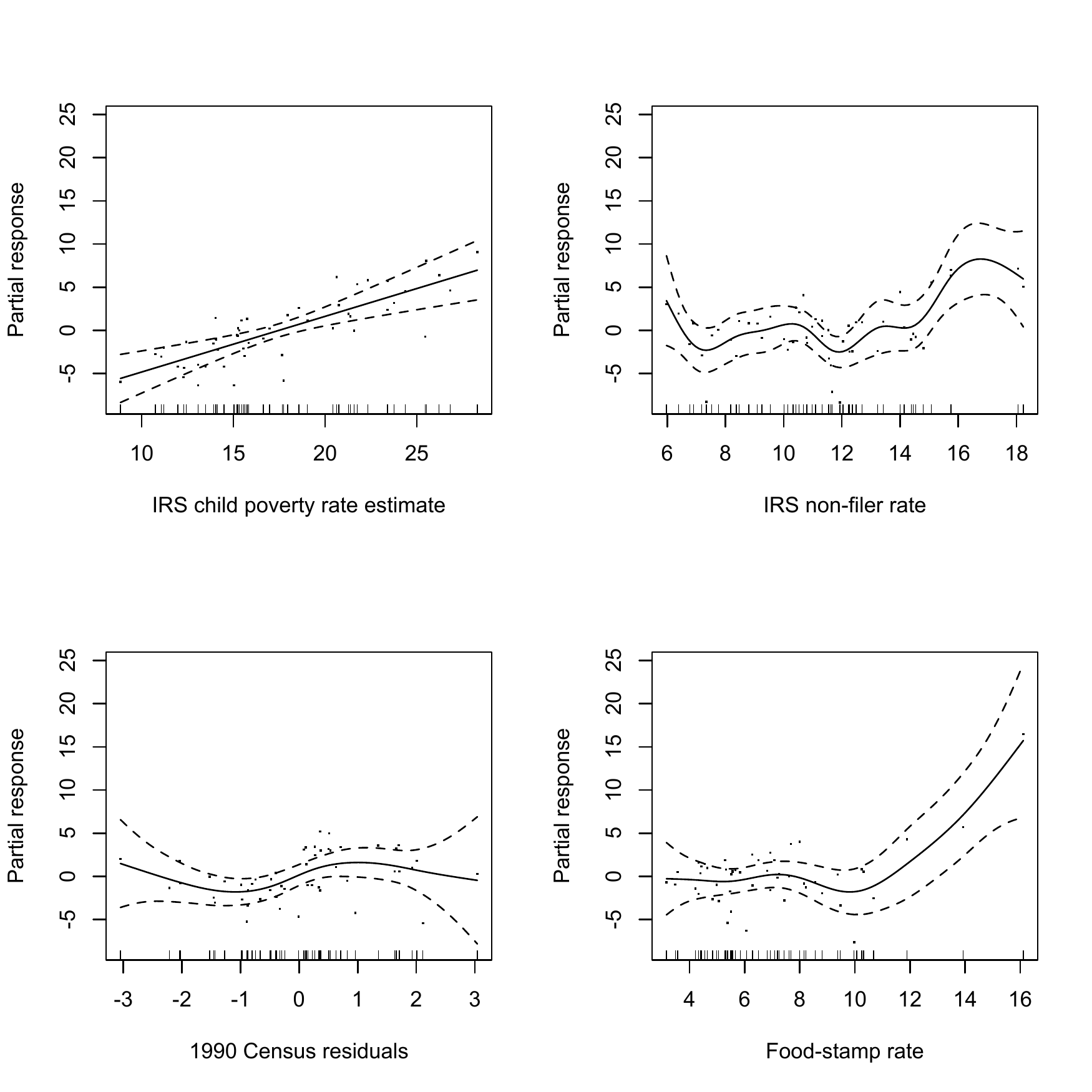}
\caption{Fitting of a generalized additive model to the 1998 SAIPE data. The
  top left plot is fairly linear, whereas the others deviate from linearity,
  suggesting the need for either a transformation or a
  nonparametric approach. In any case, this suggests that as
  in other disciplines non-linear models (e.g., generalized
    linear mixed models, kernel smoothers, splines) may be good directions to
    explore.}
\label{explore1}
\end{center}
\end{figure}

The method proposed here relies on a form of the celebrated Fay-Herriot model \citep{fay_1979}.  That model, and that of \cite{battese_1988}, were major achievements for small area estimation, but rest on simplifying assumptions, especially of linearity, which are not always valid.  These modeling assumptions were not checked in \cite{pfeffermann_2014}, and we are unable to check them ourselves since the covariates are not given.\footnote{It would be easier to perform such checks, and to make comparisons to our methods, if the code and data were made publicly available.  This has been done for small area models by \cite{molina_2013}, and advocated for statistical computation more generally by \cite{stodden_2013}.}  To illustrate what would be involved in such a check, however, we look at cross-sectional data on child poverty from the  the Small Area and Income Poverty Estimate (SAIPE) Program at the U.S. Census Bureau.\footnote{This dataset was used before in \cite{datta_2011}, with a Fay-Herriot model, and is publicly available at \url{http://www.census.gov/did/www/saipe/data/statecounty/data/1998.html}.}  The response variable is the rate of child poverty in each state (in 1998), and the covariates are an IRS income-tax-based pseudo-estimate of the child poverty rate, the IRS non-filer rate, the rate of food stamp usage, and the residual term from
the regression of the 1990 Census estimated child poverty rate.  We fit an additive model, which strictly nests the linear model implied by the Fay-Herriot assumptions.  Figure~\ref{explore1} shows the estimated partial response functions, which are clearly nonlinear for three of the four covariates.

If, on examination, the data in \cite{pfeffermann_2014} are not linear, then the exact method they propose does not apply.  However, semiparametric and non-linear models are also available in the small area literature. For example,
models including penalized splines (P-splines) have been already used. In particular, \cite{ugarte_2009} use P-splines to model and forecast dwelling prices in the different neighbourhoods  of a Spanish city and \cite{militino_2012} use P-splines to estimate the percentage of food expenditure for alternative household sizes at provincial
level in Spain using the Spanish Household Budget Survey. Generalized linear mixed models are also common in disease mapping, an important application of small area estimation. For details, see \cite{Ugarte2014} and references therein.  It seems important to determine how robust the \cite{pfeffermann_2014} method is to nonlinearity, and how it must be modified to handle highly nonlinear systems.

\section{Time Complexity, Model Comparisons, and Benchmarking}
\label{sec:bench-error}

Benchmarking requires that estimates at a lower level of aggregation (perhaps counties) must match the estimates at a level of higher aggregation (the state total for example).
\cite{pfeffermann_2013} and \cite{pfeffermann_2014} discussed many advances in small area estimation and benchmarking, making the novel contributions very clear in terms of methodological advancements.\\

\subsection{Time complexity and Model Comparisons}

The procedure of \cite{pfeffermann_2014} is not the first attempt at two-stage benchmarking for time
series.  The key difference between their method, and the earlier one of \cite{ghosh_2013}, is that the
newer procedure cannot be calculated at once, but must be computed in two separate stages.  It is not
clear to us how one ought to choose between these methods in any particular application.
A sensible reason to prefer one method would be that its underlying model fits better.  This raises issues of model specification, as we mentioned above, as well as model \emph{comparison}.  The latter is not discussed by \cite{pfeffermann_2014}, but would seem to deserve serious attention. \\

Another, less statistical criterion for choosing between methods would
computational complexity, however it is not clear which will \emph{always} be
the clear winner.  The Bayesian approach of \cite{ghosh_2013} must wait for the Markov chain Monte Carlo to come close to convergence, but once it has a posterior sample, estimation is instantaneous,
since most solutions are Bayes estimators in a reduced action space.  For
\cite{pfeffermann_2014}, on the other hand, we ask what is the computational
complexity of their method (and empirically how fast is it)? 
This requires
further analysis. 


\subsection{Benchmarking with Error}

\cite{pfeffermann_2014} clearly  explain the distinction between \emph{external} and \emph{internal} benchmarking. We focus here on external benchmarking, where there is an additional source of information that can be used for the purpose of the benchmarking procedures. Many papers have included external benchmarking in the last few years. For example,  see  \cite{datta_2011}, \cite{ghosh_2013} and  \cite{bell_2013} for details.\\

Production of small area estimation   motivates  ``traditional'' Bayesian benchmarking, wherein sub-domain estimates are optimized subject to the constraint that the weighted average equals some total for the domain (usually the direct estimate)
\citep{datta_2011,ghosh_2013, wang_2008, pfeffermann_2014}.
This is effective when, for example, the goal is to match a direct domain estimate.  However, it does not address situations where the direct estimate is imprecise,  there is additional information at the domain level (e.g., an indirect estimate), or the domain is itself a sub-domain of a higher level.  In such situations, it is more natural to treat the additional information as random rather than fixed.  Thus, instead of making it a hard external constraint, it should be embedded into a probabilistic model, along with the sub-domain data.  
We bring this up because it is hard to believe that the direct domain estimates used in \cite{pfeffermann_2014} are fixed and without error.  The uncertainty in these ``external'' quantities  ought to be incorporated into the benchmarking, using the kind of probabilistic approach just mentioned.
Preliminary investigations \citep{louis_2014} suggest that under normality and linearity, such embeddings {\em may} produce almost the same results as the ``traditional'' approach.  This equivalence breaks down as we move away from normality and linearity.  Even if the specific application of \cite{pfeffermann_2014} does prove to be nearly-linear and nearly-Gaussian, non-linearity and non-normality are important in modern applications.  We would very much like to see \cite{pfeffermann_2014} take up the challenge of probabilistic external benchmarking.

\section{Small Area in Practice}

Many users of small area estimation are more interested in practical considerations than in the underlying theory.
Given the methods of \cite{pfeffermann_2014}  for two-stage benchmarking and those of \cite{ghosh_2013},
the advantages and disadvantages of each approach are not clear except in the obvious case when the benchmarking needs to be separated into a two-step procedure due to a time-series as \cite{pfeffermann_2014} describes.
It would perhaps be helpful if the authors could provide some practical guidance as to the benefits of their proposed methodology, as well as examples of settings in which they may outperform other approaches in the literature.\\


In the simulation study, the authors considered a reasonable number of small areas and time points. However, this might not  be the case in many practical situations and typically 
out of sample areas could be present as the domains become increasing \emph{small}. Examples include certain regions, provinces, and departments of Europe. In the United States, geographical regions include states, counties, regions, blocks, block-groups, etc. Can the same type of conclusions be made if the number areas and time points are both smaller (and in the presence of out of sample areas)?
In addition, the authors give an application, dealing with states and census divisions, where the census divisions are aggregations of the states. How many of the states and census divisions are small areas after 2000 in terms of the coefficient of variation? Clarification here would be helpful towards practical applications to much smaller regions that are of interest. To summarize, do the methods always hold in practice and if not, when do they break down?\\


We again emphasize the contributions from the authors have made regarding both a review of benchmarking and  a novel extension to benchmarking in time series. These have led us  to  ask important questions surrounding model misspecification, linearity, external benchmarking, and computational costs. Are we approaching benchmarking and small area methods in the most sound way, and if not, what should change?

\paragraph{Acknowledgements} Research by R. Steorts was supported by the National Science Foundation through grants SES1130706 and DMS1043903 and research by M. Dolores Ugarte was supported by the Spanish Ministry of Science and Innovation (project MTM2011-22664 which is co-funded by FEDER grants).

\bibliography{chomp}
\label{s:biblio}

\end{document}